\documentclass[letterpaper,journal]{IEEEtran}

\usepackage[ruled,vlined]{algorithm2e}
\usepackage{amsmath}
\usepackage{amssymb}
\usepackage{booktabs}
\usepackage{bm}
\usepackage{cite}
\usepackage{cmap}
\usepackage{colortbl}
\usepackage{enumitem}
\usepackage[T1]{fontenc}
\usepackage{framed}
\usepackage{graphicx}
\usepackage[colorlinks=true,citecolor=blue,linkcolor=.,urlcolor=.]{hyperref}
\usepackage{hyperxmp}
\usepackage[misc]{ifsym}
\usepackage[utf8]{inputenc}
\usepackage{mathtools}
\usepackage{microtype}
\usepackage{multirow}
\usepackage{pbalance}
\usepackage{soul}
\usepackage{subcaption}
\usepackage{tabularx}
\usepackage{textcomp}
\usepackage{url}
\usepackage{xcolor}

\newcommand{\nf}{\bm{\mathrm{nv}}}

\DeclareMathOperator*{\argmin}{argmin}

\newcommand{\bbone}{\text{\usefont{U}{bbold}{m}{n}1}}
\MakeRobust{\bbone}

\definecolor{shadecolor}{rgb}{0.85,0.85,0.85}

\newcommand{\best}[1]{\cellcolor{gray!25}#1}
\definecolor{lightgray}{rgb}{0.875,0.875,0.875}
\sethlcolor{lightgray}

\newcommand{\mycdot}{\,\cdot\,}

\begin{document}

\title{Protecting Deep Learning Model Copyrights with Adversarial Example-Free Reuse Detection}

\author{
    \IEEEauthorblockN{Xiaokun Luan\IEEEauthorrefmark{1}, Xiyue Zhang\IEEEauthorrefmark{2}, Jingyi Wang\IEEEauthorrefmark{3}, Meng Sun\IEEEauthorrefmark{1}\textsuperscript{\textsection}}\\%
    \IEEEauthorblockA{\IEEEauthorrefmark{1}School of Mathematical Sciences, Peking University, China, \{luanxiaokun, sunm\}@pku.edu.cn}\\%
    \IEEEauthorblockA{\IEEEauthorrefmark{2}School of Computer Science, University of Bristol, UK, xiyue.zhang@bristol.ac.uk}\\%
    \IEEEauthorblockA{\IEEEauthorrefmark{3}School of Control Science and Engineering, Zhejiang University, China, wangjyee@zju.edu.cn}
}

\maketitle
\begingroup\renewcommand\thefootnote{\textsection}
\makeatletter\def\Hy@Warning#1{}\makeatother%
\footnotetext{Corresponding author}
\endgroup

\begin{abstract}
    Model reuse techniques can reduce the resource requirements for training high-performance deep neural networks (DNNs) by leveraging existing models.
    However, unauthorized reuse and replication of DNNs can lead to copyright infringement and economic loss to the model owner.
    This underscores the need to analyze the reuse relation between DNNs and develop copyright protection techniques to safeguard intellectual property rights.
    Existing DNN copyright protection approaches suffer from several inherent limitations hindering their effectiveness in practical scenarios. For instance, existing white-box fingerprinting approaches cannot address the common heterogeneous reuse case where the model architecture is changed, and DNN fingerprinting approaches heavily rely on generating adversarial examples with good transferability, which is known to be challenging in the black-box setting.
    To bridge the gap,
    we propose NFARD, a Neuron Functionality Analysis-based Reuse Detector,
    which only requires normal test samples to detect reuse relations by measuring the models' differences on a newly proposed model characterization, i.e., \emph{neuron functionality (NF)}.
    A set of NF-based distance metrics is designed to make NFARD applicable to both white-box and black-box settings.
    Moreover, we devise a linear
    transformation method to handle heterogeneous reuse cases by constructing the optimal projection matrix for dimension consistency, significantly extending the application scope of NFARD.
    To the best of our knowledge, this is the first adversarial example-free method that exploits neuron functionality for DNN copyright protection.
    As a side contribution, we constructed a reuse detection benchmark named Reuse Zoo that covers various practical reuse techniques and popular datasets.
    Extensive evaluations on this comprehensive benchmark show that NFARD achieves F1 scores of 0.984 and 1.0 for detecting reuse relationships in black-box and white-box settings, respectively, while generating test suites $2\sim 99$ times faster than previous methods.
\end{abstract}

\begin{IEEEkeywords}
    copyright protection, reuse detection, neuron functionality, deep neural networks
\end{IEEEkeywords}

\section{Introduction}
\label{sec:intro}

Deep learning (DL) has achieved remarkable success across a wide range of tasks~\cite{krizhevsky_imagenet_2017,graves_speech_2013,collobert_natural_2011}.
However, training high-performance deep neural networks (DNNs) demands vast amount of labeled data and substantial computational resources, and these costs escalate with model complexity.
Training state-of-the-art models like GPT-4~\cite{knight_openai_2023}, for instance, reportedly cost over \$100 million.
Consequently, reusing and adapting existing DL models for new tasks has become common practice to save resources.
Techniques such as transfer learning~\cite{weiss_survey_2016}, which leverages knowledge from one task for another related one, and model compression~\cite{cheng_survey_2020}, which reduces model size for deployment in resource-constrained environments, exemplify this trend towards efficient model reuse.

Despite the benefits, model reuse techniques also pose concerns regarding copyright infringement.
DNNs can be exposed in multiple ways without adequate protection, including distributed copies and Machine Learning as a Service (MLaaS)~\cite{ribeiro_mlaas_2015}.
Unauthorized reuse, such as commercialization in breach of license or theft via model extraction techniques~\cite{tramer_stealing_2016,orekondy_knockoff_2019,yuan_es_2022} (a form of model reuse), constitutes copyright infringement, posing threats to the competitive edge of the DL model owner and causing significant economic loss.
With the push for AI governance, such as the OECD recommendation on AI~\cite{oecd_recommendation_2019} and the European Union's Artificial Intelligence Act~\cite{european_proposal_2021},
DNN copyright protection is becoming increasingly important to the responsible development of DL models in both academic and industrial practices.
There is a growing need to develop DNN copyright protection methods to prevent the valuable intellectual properties from being illegally replicated and to provide the necessary evidence to resolve copyright infringement disputes.

A number of DNN copyright protection methods have been proposed, including two major types: watermarking and fingerprinting.
Watermarking methods~\cite{uchida_embedding_2017,darvish_rouhani_deepsigns_2019,wang_riga_2021,adi_turning_2018,gu_badnets_2019} embed hidden identifiers into the model's parameters during training, leveraging over-parameterization.
Then the owner can identify whether a suspect model is a copy by comparing the extracted watermarks.
However, the watermark needs to be embedded during the training phase, which may compromise the performance and security of the model.
Moreover, recent studies have shown that watermarks can be vulnerable to removal via adaptive attacks~\cite{aiken_neural_2021,shafieinejad_robustness_2021}.
In contrast, fingerprinting methods~\cite{cao_ipguard_2021,lukas_deep_2021,peng_fingerprinting_2022,chen_copy_2022,li_modeldiff_2021} do not tamper with the training procedure.
They extract fingerprints from the model using a set of carefully crafted test samples, which contains characterizing features of the model that remain unchanged after applying DNN reuse techniques.
Any model with a similar fingerprint is deemed a derivative.
However, fingerprinting methods face two main challenges.
First, most existing fingerprinting methods depend significantly on the generation of adversarial examples with high transferability, a task that is particularly difficult in black-box settings~\cite{papernot_practical_2017}.
Therefore, they mostly assume the availability of gradient information, which may not hold for deployed DL models.
Second, current fingerprinting approaches struggle to reliably detect reuse relations when the two models have different architectures or address different classification tasks (heterogeneous cases).
These limitations hinder the practical use of fingerprinting methods in diverse scenarios.

To address the aforementioned limitations, we propose a fingerprinting method called \textbf{N}euron \textbf{F}unctionality \textbf{A}nalysis-based \textbf{R}euse \textbf{D}etector (NFARD), which only requires normal test samples to detect reuse relations.
This method measures the differences between models based on a simple yet effective model characterization, i.e., \emph{neuron functionality (NF)}.
Neuron functionality has been well practiced to characterize a DNN in neural network representation studies~\cite{raghu_svcca_2017,morcos_insights_2018}.
In this work, we leverage neuron functionality for copyright protection.
Specifically, a neuron's functionality is the mapping from the input space to the one-dimensional output space of the neuron, i.e., $\mathcal{X} \rightarrow \mathbb{R}$ with $\mathcal{X}$ representing the input space.\footnote{Neuron functionality is the direct mapping function which is fundamentally different from neuron output statistics used in previous approaches.}
Recent studies~\cite{raghu_svcca_2017,morcos_insights_2018,mehrer_individual_2020} have shown that independently trained neural networks have divergent neuron functionalities.
On the other hand, neural networks obtained by reuse techniques have similar neuron functionalities to those of the original model.
Therefore, we can detect model reuse relations by measuring the similarity of neuron functionalities, which can be approximated only using normal samples, eliminating the need for adversarial example generation and its associated complexities.
A side benefit is that the accessibility of gradient information is not required, making NFARD applicable to more reuse techniques under both white-box and black-box settings.
We design specific distance metrics to measure the similarity of neuron functionality under these scenarios.
Furthermore, to tackle the challenge of heterogeneous reuse detection, NFARD incorporates a linear transformation technique to extend its application scope.
By constructing an optimal projection matrix via least squares optimization, we map potentially high-dimensional NF features from models with different architectures or tasks onto a common, comparable low-dimensional space.
This significantly extends NFARD's applicability to diverse real-world reuse scenarios where existing methods fall short.

To enable a comprehensive evaluation of NFARD, we construct a model reuse detection benchmark named Reuse Zoo, which consists of 250 neural network models, covering seven commonly used model reuse techniques, three datasets, and four popular model architectures.
We use this benchmark to evaluate NFARD and compare it with three baseline fingerprinting approaches: IPGuard~\cite{cao_ipguard_2021}, and two state-of-the-art methods, i.e., ModelDiff~\cite{li_modeldiff_2021} and DeepJudge~\cite{chen_copy_2022}.
Extensive experimental results show that NFARD can effectively detect reuse relations in both black-box and white-box settings, achieving an F1-score of 0.984 and 1.0, respectively, and is $2\sim 99$ times faster than previous methods in terms of test suite generation.

Our main contributions are:
\begin{enumerate}[leftmargin=15pt]
    \item \textbf{NFARD:} A novel, adversarial example-free DNN copyright protection method based on neuron functionality analysis, applicable in both white-box and black-box settings using normal test samples.
    \item \textbf{Heterogeneous reuse detection:} A linear transformation technique enabling NFARD to effectively handle reuse scenarios involving models with different architectures or classification tasks.
    \item \textbf{Reuse Zoo benchmark:} A publicly available benchmark with 250 models designed for evaluating DNN reuse detection methods, covering seven common reuse techniques.
    \item \textbf{Comprehensive evaluation:} Demonstrating NFARD's superior performance, broader application scope, and efficiency compared to the state-of-the-art approaches.
\end{enumerate}

\section{Background}
\label{sec:bg}

\subsection{Deep Neural Network and Neuron Functionality}

We focus on classification tasks in this work.
A deep neural network classifier is a function mapping from the input space to a probability distribution, i.e., $f:\mathcal{X}\to[0,1]^m$ where $\mathcal{X}$ is the input space and $m$ is the number of classes.
The neural network includes $L$ layers whose output dimensions are $m_1,m_2,\cdots,m_L$, respectively (with $m_L=m$).
We assume that the last layer is a fully connected layer with $m$ neurons, and its output is called the logits of the model.
The final probability distribution is obtained by applying the softmax function to the logits.
Here, $f$ defines the architecture (e.g., layer composition).
A trained model is an instance of certain architecture $f$ with a set of parameters $\theta$, denoted as $f(\mycdot;\theta)$.

In a neural network $f$, the functionality of a neuron is the mapping from the input space to the one-dimensional output space of the neuron.
For the $i$-th neuron of the $k$-th layer, its functionality is defined as $f_{i}^{(k)}: \mathcal{X}\to\mathbb{R}$.
Mathematically, $f_i^{(k)}$ is a highly non-convex function.
In practice, it can be approximated by a set of normal samples $\{\bm{x}_j\}_{j=1}^{n}$ and characterized as a vector denoted as $\nf(k,i)$:
\begin{equation}
    \nf(k,i) = (f_{i}^{(k)}(\bm{x}_1),f_{i}^{(k)}(\bm{x}_2),\cdots,f_{i}^{(k)}(\bm{x}_n))^\intercal,
    \label{eq:nv-def}
\end{equation}
We refer to the vector form neuron functionality $\nf(k,i)$ as the \emph{neuron vector}\footnote{This is also called activation vector in some other work~\cite{morcos_insights_2018}, but it is \emph{different} from layer output on a single sample.} in this work.
To the extent of our knowledge, the concept of neuron functionality was first formally introduced in \cite{gedeon_indicators_1995} to prune neural networks by identifying neurons with similar functionalities.

Recent work~\cite{raghu_svcca_2017, morcos_insights_2018} on neural network representation proposed a neuron functionality analysis approach to study DNN learning dynamics and interpretability.
They have shown that independently trained DNN models with different weight initialization have different corresponding neuron functionalities, which is in accordance with \cite{mehrer_individual_2020}'s conclusions.
That is to say, even if two DNN models $f(\mycdot;\theta_1)$ and $f(\mycdot;\theta_2)$ have the same structure, training dataset, and training setting, their corresponding neuron functionalities $\nf_1(k,i)$ and $\nf_2(k,i)$ can vary greatly if they have different weight initialization.

\subsection{Model Reuse Detection for Copyright Protection}

We aim to safeguard the copyright of DNNs by detecting model reuse relations.
Initially, the model owner trains a \emph{victim model} using his/her own resources, including the labeled data and the computational resource.
Then model reuse methods are applied to the victim model to obtain a \emph{surrogate model}, or a reused copy, which demonstrates comparable performance at a lower cost on similar or new tasks.
We consider the assessment of whether a reuse relationship constitutes an unauthorized exploit as a non-technical issue and focus on the detection of the reuse relation.

Given a victim model $f_v(\mycdot;\theta_v)$ and a \emph{suspect model} $f_s(\mycdot;\theta_s)$, the task of the reuse detector is to ascertain whether $f_s(\mycdot;\theta_s)$ is a positive suspect model, i.e., a surrogate model derived from $f_v(\mycdot;\theta_v)$ via reuse methods, or a negative suspect model, independently trained or unrelated to $f_v(\mycdot;\theta_v)$.
To facilitate this determination, the reuse detector typically employs a set of \emph{reference models} $\{f_{r_i}(\mycdot;\theta_{r_i})\}_{i=1}^{R}$~\cite{lukas_deep_2021,li_modeldiff_2021,chen_copy_2022}.
Each reference model shares the same structure as the victim model but is trained independently\footnote{In practice, hyperparameter tuning process (e.g., grid search) produces multiple independently trained models that can be used as reference models.}, providing a reference for the detector to make the decision.
This independence ensures that the reference models represent typical variations in model behavior when trained without knowledge transfer from $f_v(\mycdot;\theta_v)$.
For brevity, we will omit the parameters $\theta_v$ (and $\theta_s$, $\theta_{r_i}$) when they are clear from the context or not important, and use $f_v$ (and $f_s$, $f_{r_i}$) to represent the model.

Depending on the access of the reuse detector to the models, the detection methods can be divided into two categories: \emph{white-box} and \emph{black-box} methods.
In the white-box setting, the detector has access to the internals of the neural networks and the final probability output.
While in the black-box setting, the detector only has access to the final probability output.
In either case, we assume that a subset of the training data of the victim model is available for the reuse detector (e.g., the model owners themselves are searching for surrogate models).

Another important aspect for detection is the reuse type of the suspect model.
Depending on whether the suspect model has the same structure and the same classification task as the victim model, its reuse type can be classified as \emph{homogeneous} or \emph{heterogeneous}.
A homogeneous suspect model has the same structure as the victim model and is used for the same classification task.
On the other hand, a heterogeneous suspect model has a different structure or is used for a different classification task.
Detecting heterogeneous surrogate models is usually more challenging.
To the best of our knowledge, in previous work, only ModelDiff~\cite{li_modeldiff_2021} could handle transfer learning reuse cases, but it is not good at dealing with the other two heterogeneous reuse types.
We propose a benchmark Reuse Zoo covering  different types of surrogate models for DNN reuse detector evaluation, and we elaborate on the construction details in Section~\ref{sec:bench}.

\subsection{Model Reuse Methods}

We consider seven commonly used DNN reuse methods, including fine-tuning, retraining, pruning, quantization, knowledge distillation, transfer learning, and model extraction.

Fine-tuning and retraining techniques train the victim model for a few epochs with a relatively small learning rate, and some weight parameters may be frozen~\cite{tajbakhsh_convolutional_2016} to enable a finer control.
The difference between them is that retraining also re-initializes the weights of the last layer.
Pruning and quantization methods are compression techniques that aim to reduce DNN size by removing trivial weights or quantizing parameters to low-bit values while minimizing the drop in model performance~\cite{han_2016_deep, jacob_quantization_2018}.
Surrogate models obtained by these four methods have the same model structure and classification task as the victim model, so they are homogeneous models.
On the other hand, knowledge distillation and model extraction train surrogate models by exploiting the victim model's outputs~\cite{hinton_distilling_2015,orekondy_knockoff_2019,yuan_es_2022}, and they usually result in a different architecture; transfer learning replaces the last layer with a new one~\cite{weiss_survey_2016}, and the classification task becomes different.
Therefore, these three reuse techniques produce heterogeneous models with repsect to the victim model.

\section{Methodology}
\label{sec:meth}

This section presents NFARD, the proposed neuron functionality analysis-based reuse detection method.
We first conduct a preliminary study to demonstrate the principle of our approach.
Then we present the details of NFARD.
After that, we describe how to handle heterogeneous reuse cases by the linear transformation method.

\subsection{Insight into Reused DNN Models}

Training a DNN model involves optimizing the weights to reach a local minimum of the loss function.
Due to the non-convexity of neural networks, independently trained models may end up at different local minima with distinct weights.
In contrast, reusing a trained model to obtain a surrogate model is essentially finding another set of weights close to the original ones.
The more the reuse method leverages the victim model's knowledge, the more similar the obtained surrogate model to the victim model.
Inspired by this observation, we conduct a preliminary experiment to see if the neuron functionality of a reused model is also more similar to that of the victim model compared to independently trained reference models.

Given a set of test inputs $X=\{\bm{x}_1,\cdots,\bm{x}_n\}$, we introduce two metrics to measure the similarity of neuron functionality.
The first metric is $\mathrm{ED}^{(k)}$, which is the average Euclidean distance between the $k$-th layer's neuron vectors of two models:
\begin{equation}
    \mathrm{ED}^{(k)}(f_v,f_s)=
    \dfrac{1}{m_k}
    \sum\limits_{i=1}^{m_k}
    \lVert \nf_v(k,i)-\nf_s(k,i)\rVert _2,
\end{equation}
where $\nf_v(k,i)$ is the $i$-th neuron vector of the $k$-th layer of $f_v$, and $\nf_s(k,i)$ is that of $f_s$.
The second metric is the average cosine distance of the neuron vectors, denoted as $\mathrm{CD}^{(k)}$:
\begin{equation}
    \mathrm{CD}^{(k)}(f_v,f_s)
    = \dfrac{1}{m}
    \sum\limits_{i=1}^{m}
    (1 - \dfrac{\nf_v(k,i)^\intercal \nf_s(k,i)}{\lVert \nf_v(k,i) \rVert _2 \lVert \nf_s(k,i) \rVert _2}).
\end{equation}
Compared with $\mathrm{ED}^{(k)}$, $\mathrm{CD}^{(k)}$ focuses more on the difference in direction rather than the scale.
The smaller the metrics, the more similar the two models are in terms of neuron functionality.
We do not include metrics used in previous DNN representation studies (such as SVCCA similarity~\cite{raghu_svcca_2017}) because they target analyzing training dynamics on hidden layers where correlation analysis is essential, and our goal is to detect reuse relations using the output layer neuron vectors.

\begin{figure}[t]
    \hspace*{\fill}
    \begin{subfigure}[t]{0.48\columnwidth}
        \includegraphics[width=\linewidth]{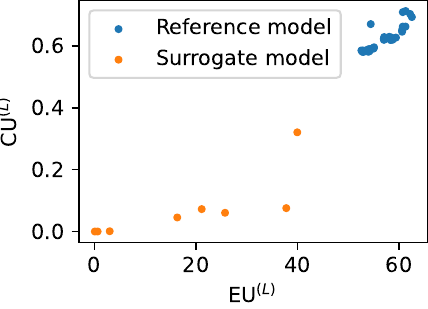}
        \caption{GoogLeNet-CIFAR10}
    \end{subfigure}
    \hfill
    \begin{subfigure}[t]{0.48\columnwidth}
        \includegraphics[width=\linewidth]{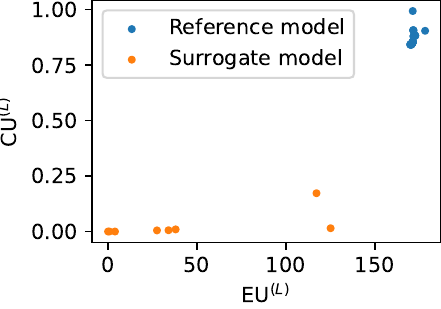}
        \caption{GoogLeNet-CIFAR100}
    \end{subfigure}
    \hspace*{\fill}

    \bigskip
    \hspace*{\fill}
    \begin{subfigure}[t]{0.48\columnwidth}
        \includegraphics[width=\linewidth]{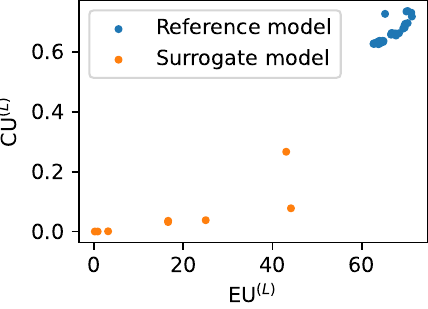}
        \caption{MobileNet V2-CIFAR10}
    \end{subfigure}
    \hfill
    \begin{subfigure}[t]{0.48\columnwidth}
        \includegraphics[width=\linewidth]{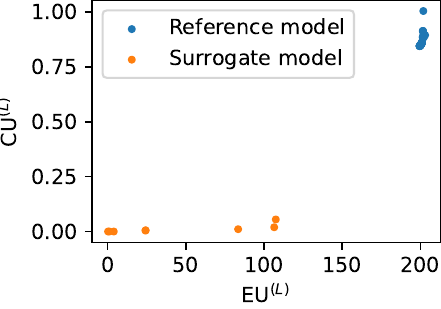}
        \caption{MobileNet V2-CIFAR100}
    \end{subfigure}
    \hspace*{\fill}
    \caption{Distance metrics between victim and suspect models.}
    \label{fig:preliminary}
\end{figure}

Next, we examine four victim models with different structures and learning tasks.
For each victim model, we prepare 8 surrogate models by applying homogeneous reuse methods and 40 reference models that are independently trained on the same task (details of these models are available in Section~\ref{sec:bench}).
We compute the distance metrics $\mathrm{ED}^{(k)}$ and $\mathrm{CD}^{(k)}$ on the last layer (before the softmax function) using 1000 normal samples, and the results are presented in Figure~\ref{fig:preliminary}, where surrogate models and reference models are plotted in orange and blue, respectively.
We can see that reference models have larger distance values and are concentrated in the upper right corner of the plane, while surrogate models have smaller distance values and are scattered on the lower left part.
Such differences indicate that homogeneous surrogate models are more similar to victim models than independently trained models regarding neuron functionality, which is exactly the principle of our proposed method.

We use the output logits to compute the metrics in this preliminary experiment.
But in practice, we can only get probability prediction obtained by applying softmax to logits.
Measuring the similarity of neuron functionality in such a black-box setting is one challenge that we need to address, and another challenge is dealing with heterogeneous reuse types.
Nevertheless, the preliminary results give us insight into the reused DNN models.

\subsection{Neuron Functionality Analysis-based Reuse Detector}

Based on the finding in our preliminary study, we propose NFARD, a neuron functionality analysis-based reuse detector.
Algorithm~\ref{alg:NFARD} demonstrates its procedure, which includes three core components: test suite selection, distance metrics, and decision criteria.
We first concentrate on dealing with homogeneous cases in white-box and black-box settings and present the linear transformation method for heterogeneous cases later.

\subsubsection{Test suite selection}

NFARD uses a subset of the victim model's training data to extract neuron vectors to compute the neuron functionality similarity between different neural networks.
Note that NFARD does not require adding any perturbations to the test samples.
To induce aligning behavior between the victim and surrogate models, we construct the test suite using samples that elicit the least confident predictions from the victim model.
The degree of confidence is measured by the entropy of the predicted probabilities.

\begin{algorithm}[t]
    \small
    \caption{Reuse detection procedure of NFARD}
    \label{alg:NFARD}
    \KwIn{victim model $f_v(\mycdot;\theta_v)$, suspect model $f_s(\mycdot;\theta_s)$, size $n$, dataset $D$, reference models $\{f_{r_i}(\mycdot;\theta_{r_i})\}_{i=1}^{R}$}
    \KwOut{decision on if $f_s(\mycdot;\theta_s)$ is a surrogate of $f_v(\mycdot;\theta_v)$}
    $X \gets TestSuiteSelection(D, n)$\;
    $distances\gets DistanceMetrics(f_v, f_s, X)$\;
    $\{ref_{i}\}_{i=1}^{R}\gets DistanceMetrics(f_v, \{f_{r_i}\}_{i=1}^{R}, X)$\;
    $decision\gets DecisionCriteria(distances, \{ref_{i}\}_{i=1}^{R})$\;
    \KwRet{$decision$}
\end{algorithm}

\subsubsection{Distance metrics}

NFARD employs different distance metrics for the white-box and black-box cases and allows extending our method by incorporating more metrics.

In the white-box setting, the detector has access to the neuron vectors of all layers, so we can directly adopt $\mathrm{ED}^{(k)}$ and $\mathrm{CD}^{(k)}$ as the distance metrics.
The parameter $k$ specifies the target layer from which we extract the neuron vectors, and the choice of target layer depends on neural network architecture.
In practice, we usually choose a relatively shallow layer as the target layer.

In the black-box setting, neither $\mathrm{ED}^{(k)}$ nor $\mathrm{CD}^{(k)}$ is directly applicable as we have the probability predictions (by applying softmax function) rather than the neuron vectors of the last layer.
It is impossible to recover the output (neuron vector) of the last layer from the probability predictions since the softmax function is not bijective.
We propose to take the logarithm of the probability to compute a logit approximation.
The differences in the computation of distance metrics in the black-box setting are shown as follows:
\begin{align}
    \mathrm{BED}(f_v,f_s) & = \dfrac{1}{m}\sum\limits_{i=1}^{m}\lVert \tilde{\nf_v}(L,i)-\tilde{\nf_s}(L,i)\rVert _2,                                                                                  \\
    \mathrm{BCD}(f_v,f_s) & = \dfrac{1}{m}\sum\limits_{i=1}^{m} (1 - \dfrac{\tilde{\nf_v}(L,i)^\intercal \tilde{\nf_s}(L,i)}{\lVert \tilde{\nf_v}(L,i) \rVert _2 \lVert \tilde{\nf_s}(L,i) \rVert _2})
\end{align}
where $\tilde{\nf}(L,i)$ is the last layer's approximated neuron vector ($f(\bm{x})_i$ is the $i$-th element of the predicted probability on $\bm{x}$):
\begin{equation}
    \tilde{\nf}(L,i) = (\log f(\bm{x}_1)_i, \log f(\bm{x}_2)_i,\cdots,\log f(\bm{x}_n)_i)^\intercal.
\end{equation}
Taking the logarithm of the probability amplifies the differences between models because the probability takes a value from 0 to 1, while its logarithm takes a value from $-\infty$ to 0.
We will demonstrate the advantage of the logarithm approximation in Section~\ref{sec:eval}.

Note that our metrics are defined on the column vectors of the neuron matrices, i.e., \textit{neuron vectors}, while the metrics of existing methods are mostly defined on the \textit{layer vectors}, usually calculated by running a single sample and collecting layer outputs.

\subsubsection{Decision criteria}

NFARD requires a set of reference models to make the final decision.
For a given distance metric $D_i$, the goal is to determine whether the distance between the victim model and the suspect model, $D_i(f_v,f_s)$, is significantly smaller than the distances between the victim model and the reference models, $\{D_i(f_v,f_{r_j})\}_{j=1}^{R}$.
To achieve this, we design a threshold function based on the interquartile range (IQR) rule, which identifies data points below the first quartile minus 1.5 times the IQR as outliers.
To balance specificity and sensitivity, we generalize this rule by introducing a sensitivity parameter $\alpha$, replacing the coefficient 1.5.
The threshold function is defined as the gap between the adjusted IQR lower bound and the distance metric value:
\begin{equation}
    \tau_i(f_v,f_s,\{f_{r_j}\}_{j=1}^{R}) = M_i - \alpha\cdot\mathrm{IQR}_i - D_i(f_v,f_s),
\end{equation}
where $M_i$ is the median of the distances $\{D_i(f_v,f_s)\}\cup\{D_i(f_v,f_{r_j})\}_{j=1}^{R}$, and $\mathrm{IQR}_i$ is their interquartile range. The user-defined parameter $\alpha$ controls the trade-off between specificity and sensitivity: higher $\alpha$ values make the criterion stricter, favoring specificity over sensitivity.
A higher value of the threshold function $\tau_i$ indicates a greater deviation of the suspect model from the reference models, thus a higher likelihood of it being a surrogate model.

Considering the varying significance and scales of different distance metrics (e.g., cosine distance is bounded, but Euclidean distance is not), the final decision is determined based on the value of the weighted sum
\begin{equation}
    \Psi(f_v,f_s,\{f_{r_j}\}_{j=1}^{R}) = \sum_{i=1}^{p}w_i\cdot \tau_i(f_v,f_s,\{f_{r_j}\}_{j=1}^{R}),
\end{equation}
referred to as the \emph{decision value},
where $w_i$ is the weight of the distance metric $D_i$ and there are $p$ distance metrics in total.
If the weighted sum is positive, NFARD identifies the suspect model as a surrogate model; otherwise, it is classified as an independently trained model.

\subsection{Handling Heterogeneous Reuse Cases}

\begin{figure}[t]
    \centering
    \includegraphics[width=0.9\linewidth]{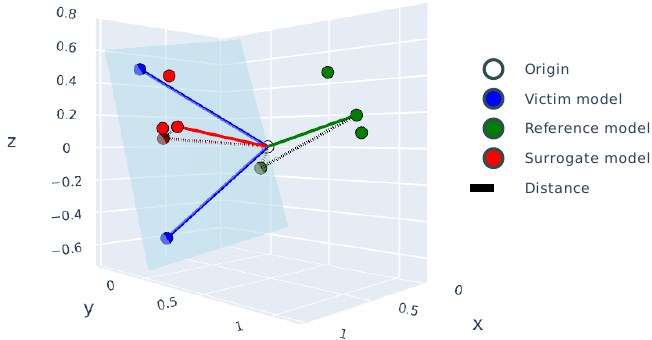}
    \caption{Illustrative example for linear transformation method.}
    \label{fig:hetero-example}
\end{figure}

The detection workflow described above only applies to \emph{homogeneous} cases where the victim and suspect models have the same classification task (in the black-box setting) or the same architecture (in the white-box setting).
However, there are also \emph{heterogeneous} cases that we need to address.

Without loss of generality, suppose we have two sets of neuron vectors from two models, $\{\nf_1(k,i)\}_{i=1}^{a}$ and $\{\nf_2(k,i)\}_{i=1}^{b}$, where $k$ is the layer index and $a\geq b$, i.e., one model has more neurons than the other.
We concatenate the neuron vectors of the two models into two matrices $\bm{H}_1\in\mathbb{R}^{n\times a}$ and $\bm{H}_2\in\mathbb{R}^{n\times b}$, where $n$ is the number of samples.
The main obstacle here is that we cannot directly compare the neuron vectors in $\bm{H}_1$ and $\bm{H}_2$ because there is no apparent functionality correspondence between the neurons of the two models with different structures.
To tackle this problem, we construct a linear transformation $\bm{P}\in\mathbb{R}^{b\times a}$ to project neuron vectors in $\bm{H}_1$ into a space of the same dimension as $\bm{H}_2$ so that the metrics we designed above become applicable to $\bm{H}_1{\bm{P}}^\intercal$ and $\bm{H}_2$.
The linear transformation $\bm{P}$ should preserve the difference among independently trained models in terms of neuron functionality, as well as the similarity between victim models and surrogate models, to enable reuse relation detection between heterogeneous models.

We construct the desired linear transformation $\bm{P}$ by minimizing the distance between neuron vectors in $\bm{H}_1$ and $\bm{H}_2$ under a linear relation, which is formulated as the following optimization problem:
\begin{equation}
    \label{eq:least-square}
    \bm{P} = \argmin_{\bm{W}}\;\lVert \bm{H}_1{\bm{W}}^\intercal-\bm{H}_2 \rVert _{F},
\end{equation}
where $\lVert\cdot\rVert _{F}$ is the Frobenius norm.
Intuitively, neuron vectors of reference models are largely linearly independent of victim model's neuron vectors since they are independently trained, so the distances between victim models and reference models will still remain relatively large after the linear transformation.

An illustrative example is provided in Figure~\ref{fig:hetero-example}.
We use colored circles to represent 3-dimensional neuron vectors.
The victim model is blue, and it has two neuron vectors.
In the heterogeneous case, three surrogate models are depicted in red, each with a single neuron vector.
Additionally, three independently trained models are colored in green, each also having one neuron vector.
The light blue plane represents the subspace spanned by the victim model's neuron vectors, and our goal is to find a vector in this plane (i.e., a linear transformation from a 2-dimensional subspace to a 1-dimensional subspace) that yields a minimum distance to a given suspect model.
This example presents the simplest case where the suspect model has only one neuron vector, and the solution to problem~(\ref{eq:least-square}) can be obtained by an orthographic projection.

As shown in this example, reference models' neuron vectors are spread further from the subspace than surrogate models' due to their lower linear correlation.
Therefore, after applying the linear transformation to the larger neuron matrix $\bm{H}_1$, we can still distinguish between independently trained models and surrogate models.
In fact, as we will show in Section~\ref{sec:eval}, distance metrics between surrogate models and victim models after applying the linear transformation method are usually about $10^{-6}\sim 10^{-5}$, while those between reference models and victim models are $1\sim 20$.

With a closer look at problem~(\ref{eq:least-square}), it solves $b$ sub-problems simultaneously, and each sub-problem can be formulated as
\begin{equation}
    \bm{p}_i = \argmin_{\bm{w}_i} \lVert \bm{H}_1 \bm{w}_i - \nf_2(k,i) \rVert _{2},
\end{equation}
where $\bm{w}_i$ is an $a$-dimensional column vector and $\bm{p}_i$ is the $i$-th row vector of $\bm{P}$.
Each sub-problem is a standard least square problem, and the closed-form solution of problem~(\ref{eq:least-square}) is
\begin{equation}
    \bm{P} = {(\bm{H}_1^\dagger\bm{H}_2)}^\intercal,
\end{equation}
where $(\cdot)^\dagger$ is the Moore-Penrose inverse, and the proof can be found in many convex optimization textbooks such as~\cite{boyd_convex_2004}.

The linear transformation method allows us to apply all metrics above to heterogeneous cases.
Specifically, for heterogeneous cases in the black-box case, the transformation is applied after taking the logarithm of the probability.
As for heterogeneous cases in the white-box case, we use the second last layer as the target layer.
This is because a DNN can be seen as a feature extractor and a linear classifier, and the output of the second last layer is generally the extracted feature vector with low dimension.
To make the comparison fair, we apply the linear transformation method to both the suspect and the reference models when the suspect is a heterogeneous model.

\section{Benchmark}
\label{sec:bench}

\begin{table*}[t]
    \caption{Complete list of models in the Reuse Zoo benchmark.}
    \label{tab:benchmark}
    \centering
    \scriptsize
    \begin{tabular}{llll}
        \toprule
        \textbf{Method}
         & \textbf{Configuration}
         & \textbf{Number}
         & \textbf{Examples}                                       \\
        \midrule
        Pre-trained
         & --
         & 4
         & \texttt{pretrain(resnet18,imagenet)}
        \\
        Transfer learning
         & Dataset: CIFAR10, CIFAR100, SVHN
         & 12
         & \texttt{transfer(cifar10,resnet50)}
        \\
        Pruning
         & Pruning ratio: 0.3, 0.6
         & 24
         & \texttt{transfer(cifar100,mobilenet\_v2)-pruning(0.6)}
        \\
        Quantization
         & Data type: 8-bit integers, 16-bit floating points
         & 24
         & \texttt{transfer(svhn,googlenet)-quantization(qint8)}
        \\
        Fine-tuning
         & Fine-tune last layer, fine-tune all layers
         & 24
         & \texttt{transfer(cifar10,resnet18)-finetune(last)}
        \\
        Retraining
         & Retrain last layer, retrain all layers
         & 24
         & \texttt{transfer(cifar10,googlenet)-retrain(last)}
        \\
        Model extraction
         & Architecture: ResNet-18, MobileNet V2, GoogLeNet
         & 9
         & \texttt{transfer(cifar100,resnet50)-stealing(resnet18)}
        \\
        Knowledge distillation
         & Architecture: ResNet-18, MobileNet V2, GoogLeNet
         & 9
         & \texttt{transfer(svhn,resnet50)-kd(googlenet)}
        \\
        Reference
         & Weight initialization
         & 120
         & \texttt{train(resnet18,cifar10,1)}
        \\
        \bottomrule
    \end{tabular}
\end{table*}

In order to evaluate our approach comprehensively, we create a benchmark named Reuse Zoo consisting of 250 models for model reuse detection.
This benchmark contains 4 pre-trained models (PTMs), 126 surrogate models, and 120 reference models with four popular classification model architectures, including ResNet-18~\cite{he_deep_2016}, ResNet-50~\cite{he_deep_2016}, MobileNet V2~\cite{sandler_mobilenetv2_2018}, and GoogLeNet~\cite{szegedy_going_2015}.

We collect four models from TorchVision~\cite{torchvision_2016} that are pre-trained on ImageNet.
For each pre-trained model, we use the transfer learning method to obtain three surrogate models trained on CIFAR10~\cite{krizhevsky_learning_2009}, CIFAR100~\cite{krizhevsky_learning_2009}, and SVHN~\cite{netzer_reading_2011}, respectively.
Each model is created by replacing the last layer with a new one and then training on the new dataset, which is the most common practice.
These 12 models constitute the transfer learning surrogate set whose source victim models are the pre-trained models.
We construct eight surrogate models for each transfer learning reuse model by applying pruning, quantization, fine-tuning, and retraining reuse methods.
Details of the construction are as follows.
\begin{enumerate}
    \item Fine-tuning. We adopt two popular fine-tuning strategies: fine-tuning only the last layer and fine-tuning all layers.
          The fine-tuning surrogate set includes 24 models.
    \item Retraining. We first re-initialize the weights of the last layer and then fine-tune it. Similar to the fine-tuning method, we retrain only the last layer and all layers to construct the retraining surrogate set with 24 models.
    \item Pruning. We first apply the global pruning method~\cite{see_2016_compression} to the victim model with the pruning ratio set as 0.3 and 0.6, respectively, and then fine-tune the pruned models. The pruning surrogate set includes 24 models.
    \item Quantization. We use 8-bit integers (\texttt{qint8}) and 16-bit floating points (\texttt{float16}) for quantization. The quantization surrogate set consists of 24 models.
\end{enumerate}

Knowledge distillation and model extraction are typically used to train smaller surrogates with different architectures from the source model.
Since the capacity of ResNet-50 is significantly larger than that of the other three architectures, we train three knowledge distillation surrogates and three model extraction surrogates (with ResNet-18, MobileNet V2, and GoogLeNet architectures) for each ResNet-50 transfer learning model.
We use the vanilla method~\cite{hinton_distilling_2015} for knowledge distillation and the Knockoff method~\cite{orekondy_knockoff_2019} for model extraction.
These models constitute the knowledge distillation surrogate set with 9 models and the model extraction surrogate set with 9 models.
The pruning, quantization, fine-tuning, retraining, knowledge distillation, and model extraction surrogate sets contain 114 surrogate models whose source victim models are the transfer learning surrogate models.
Together with the transfer learning surrogate set, we obtain a total of 126 surrogate models using these seven model reuse techniques.

We construct 120 reference models in the following way.
We independently train ten models from scratch using different weight initialization for each model architecture (four architectures) and each dataset (three datasets).
All surrogate and reference models are well-trained and have qualified performance.
The complete list of models in the Reuse Zoo benchmark is shown in Table~\ref{tab:benchmark}, where several examples of the models are also given.

This benchmark covers four popular model architectures, ranging from small networks like MobileNet V2 to large networks like ResNet-50, and various model reuse techniques, notably including heterogeneous reuse methods that can change the model structure.
We note that the authors of ModelDiff~\cite{li_modeldiff_2021} have proposed a similar DNN reuse detection benchmark referred to as ModelReuse.
Compared with ModelReuse, Reuse Zoo
(1) is constructed with three commonly used classification datasets while ModelReuse uses two fine-grained image classification (FGIC) datasets that are less representative;
(2) includes two more model architectures than ModelReuse, namely ResNet-50 and GoogLeNet, where ResNet-50 is a significantly larger model;
(3) covers more reuse techniques, such as fine-tuning and retraining on different layers, which are absent from ModelReuse;
(4) is larger, more balanced (126 surrogate models and 120 reference models vs. 84 and 28), and of higher quality (5\% accuracy gap between homogeneous models vs. 25\% in ModelReuse).

\section{Empirical Evaluation}
\label{sec:eval}

In this section, we evaluate NFARD in order to answer the following research questions:

\begin{itemize}[leftmargin=15pt]
    \item \textbf{RQ1:}\ How effective is NFARD at detecting model reuse relation, especially heterogeneous reuse cases?
    \item \textbf{RQ2:}\ Does NFARD have a broader application scope than existing DNN reuse detection methods?
    \item \textbf{RQ3:}\ How does each component affect and contribute to the performance of NFARD?
\end{itemize}

\textbf{Parameter setting.}
We implement NFARD using PyTorch~\cite{paszke_pytorch_2019} and release its source code along with the benchmark.\footnote{Artifacts available at \url{https://figshare.com/s/691d5af98dc715cf7009}}
NFARD has four parameters: the test suite size $n$, the user-specified parameter $\alpha$ for decision criteria, the distance metrics weights $\{w_i\}$, and the choice of target layer under the white-box setting.
If not specified below, we use a test suite size of $n=1000$, and we empirically set the weights of Euclidean and cosine distances as 1 and 120, respectively.
$\alpha$ is set to 0.85 for the black-box case and 3.5 for the white-box case.
The target layer is set to the layer at about 25\% depth.
We will discuss the choice of $\alpha$ and target layer later.

\textbf{Evaluation method.}
We use the Reuse Zoo benchmark to evaluate the performance of reuse detectors.
The independently trained models in the benchmark are evenly divided into two folds: the first fold serves as reference models for decision-making, and the second fold is used as a negative suspect group to evaluate the false positive rate of the reuse detectors.
We employ five reference models with the same architecture and dataset as the victim model to detect whether a suspect model is a true surrogate.
One exception is that when the victim model is pre-trained, we do not have enough models trained on ImageNet to serve as reference models (due to limited resources).
In this case, we use five reference models with the same architecture and dataset as the surrogate model.

\subsection{RQ1: Performance of NFARD}

\subsubsection{Benchmark evaluation result}

\begin{table*}[t]
    \caption{Evaluation results of NFARD under black-box and white-box settings.}
    \label{tab:nfard-results}
    \centering
    \scriptsize
    \begin{tabular}{lcccc|cccc}
        \toprule
        \multicolumn{1}{c}{}   & \multicolumn{4}{c|}{\textbf{Black-box setting}} & \multicolumn{4}{c}{\textbf{White-box setting}}                         \\
        \midrule
        \textbf{Reuse type}    &
        \textbf{Detected}      & \textbf{Decision}                               & $\mathrm{BED}$                                 & $\mathrm{BCD}$      &
        \textbf{Detected}      & \textbf{Decision}                               & $\mathrm{ED}^{(k)}$                            & $\mathrm{CD}^{(k)}$   \\
        \midrule
        Fine-tuning - last     &
        12/12                  & 49.59$\pm$7.95                                  & 3.50$\pm$0.79                                  & 0.000$\pm$0.000     &
        12/12                  & 154.68$\pm$53.3                                 & 0.00$\pm$0.00                                  & 0.000$\pm$0.000       \\
        Fine-tuning - all      &
        11/12                  & 20.98$\pm$13.3                                  & 29.6$\pm$9.20                                  & 0.022$\pm$0.014     &
        12/12                  & 150.40$\pm$54.4                                 & 3.46$\pm$1.33                                  & 0.007$\pm$0.005       \\
        Retraining - last      &
        12/12                  & 29.76$\pm$5.70                                  & 21.9$\pm$8.49                                  & 0.013$\pm$0.003     &
        12/12                  & 154.68$\pm$53.3                                 & 0.00$\pm$0.00                                  & 0.000$\pm$0.000       \\
        Retraining - all       &
        12/12                  & 10.58$\pm$4.93                                  & 38.3$\pm$6.48                                  & 0.035$\pm$0.011     &
        12/12                  & 146.25$\pm$56.2                                 & 6.15$\pm$3.00                                  & 0.019$\pm$0.013       \\
        Pruning - 0.3          &
        12/12                  & 29.62$\pm$10.9                                  & 21.6$\pm$4.30                                  & 0.015$\pm$0.010     &
        12/12                  & 149.51$\pm$54.4                                 & 4.12$\pm$1.49                                  & 0.009$\pm$0.003       \\
        Pruning - 0.6          &
        12/12                  & 24.63$\pm$9.95                                  & 25.9$\pm$4.31                                  & 0.021$\pm$0.010     &
        12/12                  & 142.85$\pm$55.5                                 & 8.13$\pm$3.39                                  & 0.031$\pm$0.014       \\
        Quantization - float16 &
        12/12                  & 52.99$\pm$8.24                                  & 0.11$\pm$0.06                                  & 0.000$\pm$0.000     &
        12/12                  & 141.86$\pm$62.8                                 & 4.98$\pm$8.67                                  & 0.065$\pm$0.113       \\
        Quantization - qint8   &
        12/12                  & 52.01$\pm$8.16                                  & 1.09$\pm$0.32                                  & 0.000$\pm$0.000     &
        12/12                  & 141.86$\pm$62.8                                 & 4.98$\pm$8.67                                  & 0.065$\pm$0.113       \\
        \midrule
        Transfer learning      &
        10/12                  & 3.618$\pm$4.06                                  & 12.2$\pm$4.44                                  & 0.006$\pm$0.005     &
        12/12                  & 142.96$\pm$83.2                                 & 16.9$\pm$12.2                                  & 0.041$\pm$0.037       \\
        Knowledge distillation &
        9/9                    & 22.92$\pm$8.49                                  & 29.0$\pm$6.87                                  & 0.025$\pm$0.006     &
        9/9                    & 84.001$\pm$10.6                                 & 0.00$\pm$0.00                                  & 0.002$\pm$0.004       \\
        Model extraction       &
        8/9                    & 12.49$\pm$10.3                                  & 37.9$\pm$13.4                                  & 0.038$\pm$0.009     &
        9/9                    & 69.243$\pm$20.9                                 & 0.00$\pm$0.00                                  & 0.125$\pm$0.151       \\
        \midrule
        \textbf{Positive}      &
        122/126                & ---                                             & ---                                            & ---                 &
        126/126                & ---                                             & ---                                            & ---                   \\
        \textbf{Negative}      &
        0/60                   & -21.74$\pm$9.20                                 & 69.0$\pm$12.8                                  & 0.056$\pm$0.017     &
        0/60                   & -68.621$\pm$35.1                                & 129.29$\pm$66.76                               & 0.956$\pm$0.124       \\
        \bottomrule
    \end{tabular}
\end{table*}

We evaluate the effectiveness of NFARD in reuse relation detection against various reuse methods under black-box and white-box settings.
The results are shown in Table~\ref{tab:nfard-results}, which includes the number of suspects classified as surrogate models, the average decision values, and the mean and standard deviation of distance metrics.
The larger the decision value, the more likely the suspect is a surrogate model.
In contrast, the larger the distance, the less likely the suspect is a surrogate model.

According to the results, NFARD achieves 100\% precision and 96.8\% ($122 / 126$) recall on this benchmark in the black-box case, and it achieves 100\% accuracy in the white-box case.
NFARD exhibits different detection results for different types of reuse methods.
The greater the modification made by the reuse method to the victim model, the greater the change in neuron functionality, and the more difficult it is for NFARD to detect.
For example, quantization and fine-tuning the last layer only slightly modify the model weights, so they are the easiest reuse types to detect and have small distance metrics and large decision values.
On the other hand, heterogeneous reuse methods change the model structure, so they are more challenging to detect in the black-box case, indicated by the smaller decision values.
Still, NFARD is capable of detecting heterogeneous reuse relations with a success rate of 90\% under the black-box setting and 100\% under the white-box setting with the help of the linear transformation method.
Clearly, surrogate models obtained by different reuse methods all have larger average decision values and smaller average distance metrics compared to negative models, which is in accordance with the conclusion we obtained in the preliminary experiment.
Compared with the black-box setting, NFARD performs better in the white-box setting: it achieves 100\% precision and 100\% recall and yields a greater gap in decision value between the positive and negative suspects.
This is as expected since we only have approximated neuron vectors in the black-box case.

\subsubsection{Comparison with baselines}

\begin{table}[t]
    \caption{Comparison with IPGuard, ModelDiff, and DeepJudge in the black-box case. The best F1-score is \hl{highlighted}.}
    \label{tab:comparison-black}
    \centering
    \scriptsize
    \begin{tabular}{lcccc}
        \toprule
        \textbf{Reuse type}    &
        \textbf{NFARD}         & \textbf{IPGuard} & \textbf{ModelDiff} & \textbf{DeepJudge} \\
        \midrule
        Fine-tuning - last     &
        12/12                  & 12/12            & 12/12              & 12/12              \\
        Fine-tuning - all      &
        11/12                  & 12/12            & 12/12              & 12/12              \\
        Retraining - last      &
        12/12                  & 12/12            & 12/12              & 12/12              \\
        Retraining - all       &
        12/12                  & 6/12             & 12/12              & 12/12              \\
        Pruning - 0.3          &
        12/12                  & 12/12            & 12/12              & 12/12              \\
        Pruning - 0.6          &
        12/12                  & 9/12             & 12/12              & 12/12              \\
        Quantization - float16 &
        12/12                  & 12/12            & 12/12              & 12/12              \\
        Quantization - qint8   &
        12/12                  & 12/12            & 12/12              & 12/12              \\
        \midrule
        Transfer learning      &
        10/12                  & 0/12             & 11/12              & 0/12               \\
        Knowledge distillation &
        9/9                    & 0/9              & 4/9                & 6/9                \\
        Model extraction       &
        8/9                    & 0/9              & 6/9                & 6/9                \\
        \midrule
        \textbf{Positive}      &
        122/126                & 87/126           & 117/126            & 108/126            \\
        \textbf{Negative}      &
        0/60                   & 0/60             & 28/60              & 1/60               \\
        \textbf{F1-score}      &
        \best{0.984}           & 0.817            & 0.863              & 0.919              \\
        \bottomrule
    \end{tabular}
\end{table}

We compare NFARD with three model reuse detectors that require generating adversarial examples, including IPGuard~\cite{cao_ipguard_2021}, ModelDiff~\cite{li_modeldiff_2021}, and DeepJudge~\cite{chen_copy_2022}.
Among them, IPGuard and ModelDiff are black-box detectors, and DeepJudge offers both black-box and white-box versions.
Therefore, we compare NFARD with these three methods in the black-box setting and only with DeepJudge in the white-box setting.

The evaluation results under different settings are shown in Table~\ref{tab:comparison-black} and Table~\ref{tab:comparison-white}, respectively.
We can see that NFARD performs slightly worse than other methods in the black-box case when detecting homogeneous reuse cases, as it fails on one case while other methods all succeed.
However, when it comes to heterogeneous reuse cases, NFARD outperforms other methods.
In the black-box case, IPGuard and DeepJudge cannot handle transfer learning reuse cases since the classification task is changed.
ModelDiff successfully detects one more transfer learning reuse case than NFARD.
However, it cannot handle knowledge distillation and model extraction cases as well as NFARD, just like the other two method.
As for the white-box setting, DeepJudge cannot handle knowledge distillation and model extraction because of different hidden layer structures.
High precision is another advantage of NFARD.
The design of the decision criteria of NFARD is more robust than other methods, and therefore NFARD is less likely to produce false alarms.
Overall, NFARD has an F1-score of 0.984 in the black-box case and 1.0 in the white-box case, both of which are higher than other DNN reuse detectors.

We also present the evaluation results of NFARD and baseline methods on the ModelReuse benchmark in Table~\ref{tab:eval-modelreuse}.
NFARD achieves a higher F1-score compared to the baseline methods, particularly excelling in heterogeneous reuse cases.
However, the reference models in the ModelReuse benchmark exhibit lower accuracy than the victim models (e.g., 55\% vs. 80\%), making them less suitable for computing reference threshold values for decision-making due to their reduced similarity to the victim models.
As a result, all methods evaluated on the ModelReuse benchmark, including NFARD, show a lower F1-score and a higher false positive rate compared to their performance on the Reuse Zoo benchmark.
Overall, the results highlight the effectiveness of NFARD in detecting model reuse relations and underscore the comprehensiveness of the Reuse Zoo benchmark relative to the ModelReuse benchmark.

\begin{table}[t]
    \caption{Comparison with DeepJudge in the white-box case. The best F1-score is \hl{highlighted}.}
    \label{tab:comparison-white}
    \centering
    \scriptsize
    \begin{tabular}{lcc}
        \toprule
        \textbf{Reuse type}    &
        \textbf{NFARD}         & \textbf{DeepJudge} \\
        \midrule
        Homogeneous            &
        96                     & 96                 \\
        \midrule
        Transfer learning      &
        12/12                  & 12/12              \\
        Knowledge distillation &
        9/9                    & 0/9                \\
        Model extraction       &
        9/9                    & 0/9                \\
        \midrule
        \textbf{Positive}      &
        126/126                & 108/126            \\
        \textbf{Negative}      &
        0/60                   & 0/60               \\
        \textbf{F1-score}      &
        \best{1.0}             & 0.919              \\
        \bottomrule
    \end{tabular}
\end{table}

\begin{table}[t]
    \caption{Comparison of NFARD with baseline on the ModelReuse benchmark. The best F1-score is \hl{highlighted}.}
    \label{tab:eval-modelreuse}
    \centering
    \scriptsize
    \begin{tabular}{lcccc}
        \toprule
        \textbf{Reuse type}    & \textbf{NFARD} & \textbf{IPGuard} & \textbf{ModelDiff} & \textbf{DeepJudge} \\
        \midrule
        Pruning - 0.2          & 12/12          & 10/12            & 12/12              & 12/12              \\
        Pruning - 0.5          & 12/12          & 9/12             & 12/12              & 12/12              \\
        Pruning - 0.8          & 6/12           & 2/12             & 9/12               & 12/12              \\
        Quantization - qint8   & 12/12          & 12/12            & 12/12              & 12/12              \\
        \midrule
        Transfer learning      & 9/12           & 0/12             & 10/12              & 0/12               \\
        Knowledge distillation & 10/12          & 0/12             & 9/12               & 12/12              \\
        Model stealing         & 12/12          & 0/12             & 2/12               & 0/12               \\
        \midrule
        \textbf{Positive}      & 73/84          & 33/84            & 66/84              & 60/84              \\
        \textbf{Negative}      & 3/28           & 0/28             & 4/28               & 0/28               \\
        \textbf{F1-score}      & \hl{0.913}     & 0.564            & 0.857              & 0.833              \\
        \bottomrule
    \end{tabular}
\end{table}

\begin{snugshade}
    \noindent\textbf{Answer to RQ1:}
    NFARD achieves an F1-score of 0.984 in the black-box case and an F1-score of 1.0 in the white-box case on the Reuse Zoo benchmark, outperforming previous methods.
    Besides, NFARD performs well on various heterogeneous reuse cases, which many baseline methods fail to detect.
\end{snugshade}

\subsection{RQ2: Application Scope}

\begin{table}[t]
    \caption{Qualitative comparison of the application scope of NFARD with adversarial example-based methods.}
    \label{tab:scope-compare}
    \centering
    \scriptsize
    \begin{tabular}{lcccc}
        \toprule
        \textbf{Factor}     & \textbf{NFARD} & \textbf{IPGuard} & \textbf{ModelDiff} & \textbf{DeepJudge} \\
        \midrule
        Heterogeneous cases & Yes            & No               & Partial            & Partial            \\
        Fully black-box     & Yes            & No               & No                 & No                 \\
        Require gradients   & No             & Yes              & Yes                & Yes                \\
        Efficiency          & High           & Low              & Medium             & Medium             \\
        Require labels      & No             & Yes              & No                 & Yes                \\
        \bottomrule
    \end{tabular}
\end{table}

\begin{table}[t]
    \caption{Test suite generation time (in seconds) of reuse detectors. The shortest generation time is \hl{highlighted}.}
    \label{tab:time}
    \centering
    \scriptsize
    \begin{tabular}{lcccc}
        \toprule
        \textbf{Detector}  &
        \textbf{ResNet-18} & \textbf{ResNet-50} & \textbf{MobileNetV2} & \textbf{GoogLeNet} \\
        \midrule
        NFARD              &
        \best{25.2}        & \best{69.2}        & \best{27.8}          & \best{28.5}        \\
        IPGuard            &
        846.7              & 4405.2             & 1831.1               & 2826.2             \\
        ModelDiff          &
        69.5               & 198.6              & 80.3                 & 89.1               \\
        DeepJudge-black    &
        60.8               & 152.1              & 72.4                 & 77.8               \\
        DeepJudge-white    &
        141.4              & 488.5              & 2064.0               & 306.4              \\
        \bottomrule
    \end{tabular}
\end{table}

We have shown that NFARD is capable of detecting more reuse types than other DNN reuse detection methods by employing the linear transformation method.
To answer the second research question, we further compare the application scope of NFARD with other methods in the following aspects: model access, scalability, and label information.
We provide a brief summary in Table~\ref{tab:scope-compare} and describe them in detail.

\subsubsection{Model access}

NFARD requires less model access compared with adversarial example-based DNN detection methods.
In the black-box case, NFARD works in a fully black-box manner, while adversarial example-based DNN reuse detectors need gradients of the victim model to generate test samples.
For models already deployed on edge devices, even the model owner may not have access to the gradients, which makes the adversarial example-based methods inapplicable.
Although ModelDiff also has a fully black-box version, its performance severely degrades, and the authors consider it immature.
It is possible to detect DNN reuse relations by generating adversarial examples in a fully black-box manner, but it would lead to a higher cost of the generation process.

\subsubsection{Scalability}

Another important factor that affects the application scope is scalability, which is mainly limited by the efficiency of test suite generation as the metric computation and decision-making process usually takes no more than a few seconds.
We report the average test suite generation time for different model architectures of NFARD and other adversarial example-based methods (with $n=1000$) in Table~\ref{tab:time}.
The test suite generation process for NFARD is the same in black-box and white-box cases, while DeepJudge takes two different procedures for test suite generation in each case.
NFARD is $2\sim 99$ times faster than the baseline reuse detectors, and its efficiency is also less influenced by the complexity of the model.
Since NFARD directly selects samples from the training dataset, the test suite generation time stays almost the same as the number of test cases grows.
On the contrary, the test suite generation time of adversarial example-based reuse detectors grows linearly with the number of test cases.

\subsubsection{Data label}

We remark that one of the state-of-the-art methods DeepJudge, requires labeled samples to compute the similarity metrics, and IPGuard needs label information to generate test samples near the decision boundary.
This is not an unusual requirement since we also assumed that a subset of training samples is available for detection.
However, in some rare situations where normal samples are not labeled, such a requirement may limit the applicability of the detection method.
In contrast, NFARD and ModelDiff need no label information throughout the detection process, making them slightly more suitable for rare cases than the above methods.

\begin{snugshade}
    \noindent\textbf{Answer to RQ2:}
    NFARD requires no gradients of the victim model and no label information from the normal samples, and it is $2\sim 99$ times faster than adversarial example-based methods, thus having better scalability.
    Therefore, NFARD has a broader scope of application than previous methods.
\end{snugshade}

\subsection{RQ3: Influencing Factors}

Next, we explore how different configurations affect the performance of NFARD.
First, we perform an ablation study to evaluate the effectiveness of the logarithm approximation, as well as the necessity of the distance metrics.
Then, we test the performance with different parameters, including decision parameter $\alpha$, target layer choice, and test suite size $n$.

\subsubsection{Logarithm approximation and distance metrics}

\begin{table}[t]
    \caption{Evaluation results of NFARD without taking the logarithm under the black-box setting.}
    \label{tab:log-approx}
    \centering
    \scriptsize
    \begin{tabular}{lcccc}
        \toprule
        \textbf{Reuse type}
         & \textbf{Detected} & \textbf{Decision} & $\mathrm{BED}$ & $\mathrm{BCD}$ \\
        \midrule
        Fine-tuning - last
         & 12/12             & 44.61$\pm$11.0    & 0.26$\pm$0.15  & 0.001$\pm$0.0
        \\
        Fine-tuning - all
         & 10/12             & 15.08$\pm$13.6    & 3.04$\pm$1.76  & 0.23$\pm$0.08
        \\
        Retraining - last
         & 12/12             & 32.03$\pm$8.57    & 1.28$\pm$0.62  & 0.10$\pm$0.05
        \\
        Retraining - all
         & 11/12             & 7.772$\pm$6.78    & 3.26$\pm$1.67  & 0.28$\pm$0.07
        \\
        Pruning - 0.3
         & 12/12             & 25.76$\pm$12.1    & 2.13$\pm$1.20  & 0.14$\pm$0.04
        \\
        Pruning - 0.6
         & 12/12             & 22.57$\pm$9.06    & 2.11$\pm$1.07  & 0.17$\pm$0.05
        \\
        Quantization - float16
         & 12/12             & 44.98$\pm$10.9    & 0.02$\pm$0.01  & 0.0$\pm$0.0
        \\
        Quantization - qint8
         & 12/12             & 44.87$\pm$11.0    & 0.09$\pm$0.05  & 0.0$\pm$0.0
        \\
        \midrule
        Transfer learning
         & 10/12             & 0.361$\pm$0.40    & 0.31$\pm$0.24  & 0.01$\pm$0.01
        \\
        Knowledge distillation
         & 6/9               & 8.158$\pm$14.1    & 2.59$\pm$1.34  & 0.22$\pm$0.08
        \\
        Model extraction
         & 4/9               & 2.403$\pm$15.4    & 2.66$\pm$1.34  & 0.27$\pm$0.10
        \\
        \midrule
        \textbf{Positive}
         & 113/126           & ---               & ---            & ---
        \\
        \textbf{Negative}
         & 4/60              & -12.57$\pm$6.03   & 5.21$\pm$2.28  & 0.44$\pm$0.11
        \\
        \bottomrule
    \end{tabular}
\end{table}

We report the evaluation results of NFARD without taking the logarithm in the black-box setting in Table~\ref{tab:log-approx}.
There is a significant drop in performance without the logarithm approximation, especially for knowledge distillation and model extraction reuse cases.
The decision value also decreases for most reuse cases, suggesting that the logarithm approximation method does amplify the difference between independently trained models and surrogate models in the black-box case.
The F1-score of NFARD using only the BED metric or the BCD metric under the black-box setting is 0.9756 and 0.9520, respectively, which is lower than the F1-score of 0.984 using both metrics.
This indicates that the distance metrics are complementary to each other and contribute to the overall performance.

\subsubsection{Decision parameter}

\begin{figure}[t]
    \hspace*{\fill}
    \begin{subfigure}[t]{0.48\columnwidth}
        \includegraphics[width=\linewidth]{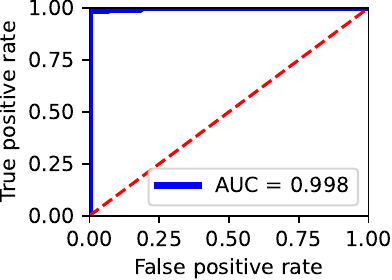}
        \caption{Black-box}
    \end{subfigure}
    \hfill
    \begin{subfigure}[t]{0.48\columnwidth}
        \includegraphics[width=\linewidth]{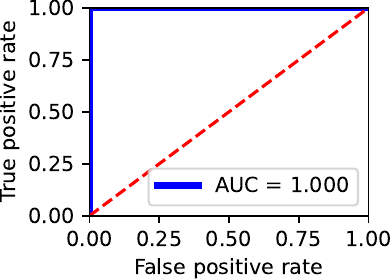}
        \caption{White-box}
    \end{subfigure}
    \hspace*{\fill}
    \caption{ROC curve of NFARD with varying decision parameter $\alpha$ under black-box and white-box settings.}
    \label{fig:alpha-roc}
\end{figure}

The decision parameter $\alpha$ balances sensitivity and specificity of NFARD.
A higher $\alpha$ value yields more conservative detection, reducing false alarms but also recall rates.
We show the ROC curve of NFARD with varying $\alpha$ under black-box and white-box settings in Figure~\ref{fig:alpha-roc}.
The area under the ROC curve for the two cases is 0.998 and 1.0, respectively.
Note that the ROC curves are plotted by varying $\alpha$ instead of directly changing the decision thresholds.
The evaluation indicates that the performance of NFARD is quite robust to the choice of $\alpha$.
We recommend using $\alpha<1$ for the black-box case and $\alpha>2$ for the white-box case in practice.

\subsubsection{Target layer}

\begin{figure}[t]
    \centering
    \includegraphics[width=0.9\columnwidth]{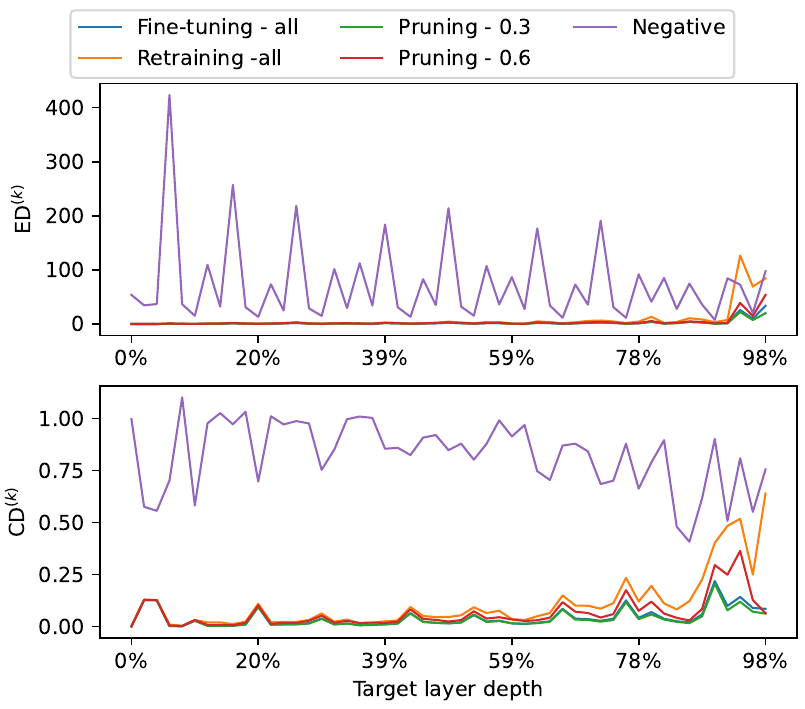}
    \caption{Distance metrics of different reuse cases under the white-box setting with target layer at different depths.}
    \label{fig:layer-effect}
\end{figure}

We evaluate how the choice of target layer affects the performance of NFARD in the white-box case by running the detection on ResNet-18 victim models with the target layer at different depths.
Figure~\ref{fig:layer-effect} shows that the distance metric gap between surrogate models and independently trained reference models tends to decrease as the target layer goes deeper.
This finding is consistent with the suggestion of choosing a shallow layer as the target layer in \cite{chen_copy_2022}.
There are two possible reasons:
one is that shallow layers learn low-level features, which tend to be similar despite being modified by reuse methods;
the other is that the model deviation accumulates as the target layer goes deeper.

\subsubsection{Test suite size}

\begin{figure}[t]
    \centering
    \includegraphics[width=0.9\columnwidth]{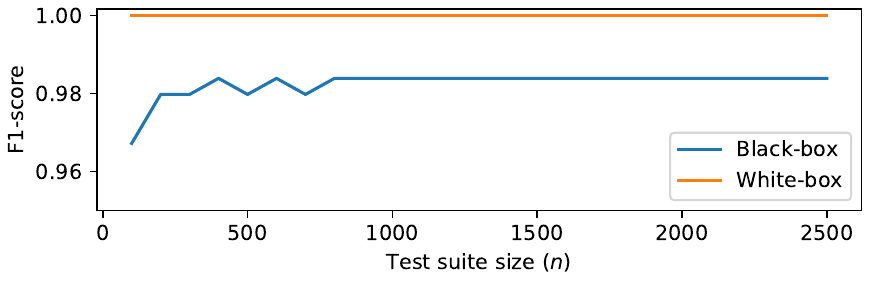}
    \caption{F1-score of NFARD under black-box and white-box settings with different test suite size.}
    \label{fig:size-effect}
\end{figure}

We evaluate NFARD on Reuse Zoo with varying test suite sizes under black-box and white-box settings and report the detector F1-score in Figure~\ref{fig:size-effect}.
We can see that NFARD achieves a perfect 1.0 F1-score under the white-box setting with different test suite sizes, indicating that 100 test samples are sufficient for effective detection in the white-box case.
On the other hand, the F1-score curve of NFARD under the black-box setting increases as the test suite size increases until it reaches a plateau with the value of $n$ around 800.
In our experiments, we set the test suite size $n$ as 1000 to balance the trade-off between effectiveness and efficiency.

\begin{snugshade}
    \noindent\textbf{Answer to RQ3:}
    Logarithm approximation contributes to the effectiveness of NFARD in the black-box case.
    Using a shallow layer as the target layer performs better in the white-box case.
    NFARD is quite robust to decision parameter $\alpha$ under both circumstances, and we recommend using 1000 test samples in practice.
\end{snugshade}

\section{Discussion}
\label{sec:discuss}

\subsection{Threats to Validity}

Both the proposed method NFARD and the baseline methods have several parameters that affect the performance, and this may affect the validity of the evaluation conclusions if these parameters are not set properly.
To mitigate this, on the one hand, we follow the recommended parameter settings by the authors of the baseline methods to ensure their optimal performance.
On the other hand, we carry out a series of experiments to explore the influence of NFARD parameters and provide the recommended settings.

The quality of the proposed benchmark is another potential threat.
Unbalanced reuse samples, limited kinds of reuse techniques, and unqualified DNN models could put the validity of conclusions at risk.
As mitigation, firstly, we include many independently trained models in the benchmark to balance the ratio of positive and negative samples (126: 120).
Secondly, we apply seven commonly used reuse techniques with different hyperparameters (e.g., global pruning with varying pruning ratios) to construct the benchmark.
Lastly, we also refer to some of PyTorch's official training recipes~\cite{vryniotis_how_2021} to ensure all trained models have qualified performance.
In addition, we release the source code for building the benchmark and binary models for future study.

The Reuse Zoo benchmark mainly focuses on image classification tasks and convolutional neural networks, and other learning tasks (e.g., text classification) and model structures (e.g., RNNs and transformers) are not included in our evaluation.
Therefore, the conclusions obtained from our experiments may not generalize to other learning tasks and DNN structures.
Most existing studies on DNN copyright protection face such risk, except that \cite{chen_copy_2022} also evaluated their method on a simple LSTM model trained for an audio classification task.
We believe that assessing the performance of copyright protection methods in more diverse scenarios could better demonstrate their usability, and we leave this as future work.

\subsection{Comparison with Existing Approaches}

Here we provide a detailed description of the two state-of-the-art methods, i.e., DeepJudge and ModelDiff, and compare them with NFARD.

DeepJudge first generates a set of adversarial examples for the victim model, then uses these adversarial examples to evaluate the similarity between the victim model and the suspect model through a set of distance metrics, e.g., the difference between the numbers of correctly classified adversarial examples by the two models (referred to as \emph{robustness distance}).
Subsequently, the final decision is determined by comparing the similarity scores against thresholds derived from reference models.
The choice of distance metrics for measuring similarity varies between the white-box and black-box settings, as certain metrics rely on the hidden layer outputs of the victim model.
Nevertheless, the rationale behind DeepJudge is to exploit the transferability of adversarial examples to detect reuse relationships, which is fundamentally different from our approach.
Therefore, DeepJudge requires gradients of the victim model to efficiently generate adversarial examples.
It also needs ground truth labels to compute the distance metrics.
In addition, DeepJudge cannot handle surrogate models obtained by transfer learning under the black-box setting due to the change of classification task.

Modeldiff follows a similar pipeline to DeepJudge, but differs in the adversarial example generation process and the similarity metrics.
ModelDiff generates adversarial examples by maximizing a customized score function, which is designed to ensure a high divergence between the original outputs and the adversarial outputs, and at the same time to increase their diversity.
The similarity metric in ModelDiff is the cosine similarity between the two models' \emph{decision distance vector}s.
A decision distance vector (DDV) of a model $f$ is a vector composed of cosine distances between $f$'s outputs on normal samples and adversarial samples.
Therefore, ModelDiff also requires gradients to generate adversarial examples.
However, it does not need ground truth labels since its similarity metric is task-agnostic, and it does not differentiate between the white-box and black-box settings.
ModelDiff can handle heterogeneous suspect models, but it struggles to effectively detect knowledge distillation and model extraction cases.

In contrast, NFARD is principled on the neural functionality differences among independently trained models, thus it does not require adversarial examples.
This feature makes it more efficient and applicable to fully black-box settings where the victim model's gradients are not accessible.
Moreover, NFARD is designed to handle various heterogeneous reuse types, and the experimental results show that it performs better at handling heterogeneous reuse cases than existing methods.
The no need for ground truth labels is another minor advantage of NFARD, which makes it more general and applicable to various tasks.

The proposed method still has several limitations.
Firstly, although it is applicable to both black-box and white-box settings, the detection accuracy in the black-box setting is lower than that in the white-box setting.
This is because lacking the victim model's internal information makes it more challenging to detect reuse relationships.
Secondly, the selection of reference models might pose bias to the detection results.
To mitigate this, one could use more reference models, but this would result in a higher computational cost, especially when the victim model is large.
Thirdly, our method requires a representative dataset for model characterization, which may not be available, or the number of samples may be limited.

\subsection{Generalization to Other Scenarios}

The black-box and white-box settings are two common scenarios in DNN reuse detection, in which cases our proposed method has shown good performance.
Its applicability to other scenarios, such as distributed learning, remains to be explored.
Nevertheless, NFARD is the first method (to the best of our knowledge) that can operate without access to the internals of the victim model, which is a significant step towards practical reuse detection.

We considered seven commonly used reuse techniques in this work, but there are other model modification techniques that could be used to derive surrogate models, such as sparse coding.
While these techniques are not covered in this work, the proposed method should remain effective in detecting reuse relationships in these cases, as long as the modified model is intended to retain the victim model's functionality.
In fact, the better the surrogate model preserves the victim model's behavior, the easier it is for NFARD to detect the reuse relationship.

\section{Related Work}
\label{sec:related}

\subsection{DNN Watermarking and Fingerprinting}

DNN intellectual property protection can be divided into two categories: active protection and passive protection.
Active protection prevents IP infringement by managing the authorized usage or perturbing the inference results, while passive protection aims to enable IP violation detection.
Here we focus on passive protection, which includes DNN watermarking and DNN fingerprinting.

Watermarking methods embed owner-specific information into DNN for IP violation detection~\cite{fkirin_copyright_2022}.
White-box watermarking methods~\cite{uchida_embedding_2017,darvish_rouhani_deepsigns_2019,wang_riga_2021} exploit the over-parameterization property of neural networks to plant secret signatures into the model weights through regularization during training.
The ownership can be checked by extracting the signatures from the model weights.
Black-box watermarking methods~\cite{adi_turning_2018,gu_badnets_2019} use a backdoor attack to implant a watermark in the model during training.
The watermark can be extracted by querying the model with triggering samples.
Either way, watermarking methods need to be applied in the training process, which will affect the model performance, and there are also methods~\cite{aiken_neural_2021,shafieinejad_robustness_2021} proposed for removing the watermarks.

Fingerprinting methods are post hoc detection techniques that do not tamper with the training process.
They extract characterizing features (called fingerprints) from the models using carefully crafted input samples close to the decision boundary~\cite{cao_ipguard_2021,lukas_deep_2021}.
A recent study~\cite{peng_fingerprinting_2022} utilizes universal adversarial perturbations~\cite{moosavi_universal_2017} to achieve similar effects.
ModelDiff~\cite{li_modeldiff_2021} and DeepJudge~\cite{chen_copy_2022} are the two state-of-the-art fingerprinting methods.
ModelDiff can handle the transfer learning reuse case, and DeepJudge has a white-box detection procedure and achieves a higher detection accuracy on homogeneous reuse cases.
Besides, \cite{chen_fedright_2023} proposes a fingerprinting method for protecting copyrights of federated learning models.

\subsection{Active DNN Copyright Protection}

Different from passive protection methods, active protection aims to prevent copyright infringement from occurring.
For distributed model copies, DNN authentication techniques~\cite{chen_deepattest_2019,chakraborty_hardware_2020,lin_chaotic_2021} manage the authorized usage of the source model by encoding the model with an authentication key, so that only authorized users with valid keys can access the model normally, while adversaries with invalid keys can only use the model with greatly degraded performance, thus preventing unauthorized model usage.
As for MLaaS mode, model extraction is the major threat, and inference perturbation techniques~\cite{jutti_prada_2019,lee_defending_2019,kariyappa_defending_2020} protect the source model by modifying the inference results before returning them to the user.

\subsection{Neural Network Representation}

Our approach is inspired by the line of research on neural network representation.
Canonical correlation analysis techniques such as SVCCA~\cite{raghu_svcca_2017} and PWCCA~\cite{morcos_insights_2018} are proposed to analyze the representation of deep neural networks.
These analysis techniques contribute to understanding the interpretability of neural networks and learning dynamics.
Previous studies concentrate on revealing the individual differences among independently trained models~\cite{mehrer_individual_2020} and the properties of converged weights of neural networks~\cite{morcos_insights_2018} by extensive experiments.
The results indicate that independently trained neural networks usually have divergent neuron functionalities, even though they have the same architecture and are trained with the same dataset.
By exploiting the fact that models obtained by reuse techniques have similar neuron functionalities, we devise our adversarial example-free detection method.

\section{Conclusion}
\label{sec:conclude}

In this paper, we presented NFARD, a novel deep neural network copyright protection method centered on neuron functionality analysis.
The core innovation lies in circumventing the need for adversarial examples, enabling robust model reuse detection using only normal test samples.
This makes NFARD particularly well-suited for practical black-box deployments.
Additionally, our proposed linear transformation technique effectively tackles heterogeneous reuse scenarios across different architectures and taskss, significantly extending its application scope.
Evaluation results demonstrate NFARD's effectiveness and efficiency, offering a versatile tool for detecting model reuse and safeguarding intellectual property rights against unauthorized replication.

\section*{Acknowledgment}
\addcontentsline{toc}{section}{Acknowledgment}

This research was sponsored by National Natural Science Foundation of China under Grant No. 62172019 and National Key R\&D Program of China under Grant 2022YFB2702200.

\bibliographystyle{IEEEtran}
\bibliography{TNNLS-2024-P-36308.R1}

\end{document}